\documentclass{article}

\usepackage{arxiv}
\usepackage{amsmath,amssymb,amsfonts}
\usepackage{hyperref}
\usepackage{academicons}
\usepackage{url}            
\usepackage{graphicx}
\usepackage{xcolor}
\usepackage{colortbl}
\usepackage{lipsum}
\usepackage{tabularx}
\usepackage{booktabs}
\usepackage{makecell}
\usepackage[tableposition=top]{caption} 
\usepackage[noend]{algpseudocode}
\usepackage[linesnumbered,ruled]{algorithm2e}

\title{Cross-Language Source Code Clone Detection Using Deep Learning with InferCode}

\author{
	\href{https://orcid.org/0000-0001-9686-3385}{\includegraphics[scale=0.06]{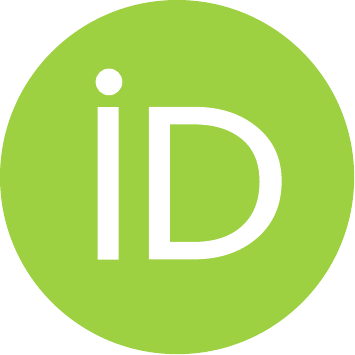}\hspace{1mm}Mohammad A. Yahya} \\
	Department of Computer Science\\
	Oakland University\\
	Rochester, Michigan \\
	\texttt{yahya@oakland.edu} \\
	\And
	\href{https://orcid.org/0000-0002-7133-9111}{\includegraphics[scale=0.06]{orcid.pdf}\hspace{1mm}Dae-Kyoo Kim} \\
	Department of Computer Science\\
	Oakland University\\
	Rochester, Michigan \\
	\texttt{kim2@oakland.edu} \\
}

\hypersetup{
	pdftitle={A template for the arxiv style},
	pdfsubject={q-bio.NC, q-bio.QM},
	pdfauthor={Mohammad A. Yahya, Dae-Kyoo Kim},
	pdfkeywords={Cross-Language Source Code Clone Detection Using Deep Learning with InferCode},
}

\begin{document}
\maketitle
\begin{abstract}
Software clones are beneficial to detect security gaps and software maintenance in one programming language or across multiple languages. The existing work on source clone detection performs well but in a single programming language. However, if a piece of code with the same functionality is written in different programming languages, detecting it is harder as different programming languages have a different lexical structure. Moreover, most existing work rely on manual feature engineering. In this paper, we propose a deep neural network model based on source code AST embeddings to detect cross-language clones in an end-to-end fashion of the source code without the need of the manual process to pinpoint similar features across different programming languages. To overcome data shortage and reduce overfitting, a Siamese architecture is employed. The design methodology of our model is twofold -- (a) it accepts AST embeddings as input for two different programming languages, and (b) it uses a deep neural network to learn abstract features from these embeddings to improve the accuracy of cross-language clone detection. The early evaluation of the model observes an average precision, recall and F-measure score of $0.99$, $0.59$ and $0.80$ respectively, which indicates that our model outperforms all available models in cross-language clone detection.
\end{abstract}
	

\section{Introduction}
\label{sec:intro}
Code clone (CL) detection is the process of detecting similar code fragments. Some code fragments are copied online without realizing potential negative effects (e.g., security threats, increasing complexity). Detecting such clones is easier if the code snippets are in the same programming language~\cite{saini2018oreo, sajnani2016sourcerercc, kamiya2002ccfinder, perez2019cross, nafi2019clcdsa, wu2020deep}.
Duplicate functionalities are common in a large software system where the software is often written in multiple programming languages. If one functionality is to be updated or removed, this has to be reflected on all clones. On one hand, the latest approach by Perez and Chiba~\cite{perez2019cross} on cross-language clones detection relies on skip-gram models over AST. However, it ignores the morphology of tokens, which impairs the accuracy of detection. Furthermore, it is not clear how their model can capture the trees that greatly differ syntactically. For example, although the ASTs of the code snippets in Fig.~\ref{fig:java_py_clone} look completely different as they are written in different programming languages, they have the same functionality. So it becomes difficult for an AST-based model to recognize such a clone as reported in their evaluation. This shows that a careful selection of source code embedding technique is critical for accuracy as we will discuss how different embeddings affect F1 score.

Detecting code clones has long been an active area of research~\cite{baker1993program} with various approaches proposed including token, AST, metrics, binary and graph-based approaches~\cite{baxter1998clone, kamiya2002ccfinder, di2002approach, saebjornsen2009detecting, krinke2001identifying,roy2018benchmarks}. Most existing work focuses on clone detection in the same programming language. However, common APIs (e.g., Apache Spark) for big data processing have similar naming and call patterns written for different programming languages~\cite{nafi2019clcdsa, perez2019cross}. Not all code bases use similar APIs. To date, the latest approaches are not end-to-end, meaning that exhaustive feature engineering has to take place to prepare data. Our preliminary study~\cite{lei2022deep} shows that about 90\% of code clone detection using deep learning was performed on top of a single language and only \%10 was about cross-language clone detection. The latest approaches on cross-language code clone detection use low quality features for training, which makes it difficult to train a Siamese architecture which we use in our work to improve accuracy. Code clone fits into four main categories-- Type I which is textual clones (i.e., identical code), Type II is lexical clones (e.g., different identifiers), Type III which is syntactic clones (e.g., additional statements), and Type IV which is semantic clones or purely semantic (i.e., the same logic)~\cite{bellon2007comparison, roy2007survey}. In this work, we focus on Type III. It has been reported that Type III clones are more frequent than other types of clones~\cite{svajlenko2014towards, roy2010near, roy2014vision}.

In this paper, we present an end-to-end cross-language clone detection approach based on embeddings of ASTs using InferCode~\cite{bui2021infercode}. Inspired by TBCNN~\cite{mou2016convolutional, mou2015discriminative}, InferCode captures the syntactical essence of a program as a vector. Our model takes the pre-trained embeddings from InferCode as input to learn abstract features by grouping related embeddings through the contrastive loss function which guarantees to have similar embeddings together, but dissimilar embeddings far apart. We make the following contributions.

\begin{itemize}
\item We present a cross-language clone detection across any two languages. In this work, we demonstrate finding clones between Java and Python source code.  
\item We created a dataset of pairs containing around 40,000 code snippets that contain code that has similar and dissimilar pairs. 
\item We improved the state-of-the-art work for cross-language code clone detection by over \%10. 
\end{itemize}

The remainder of the paper is organized as follows. Section~\ref{sec:related} covers up-to-date approaches in clone detection with their pros and cons. Section~\ref{sec:background}  discusses InferCode in comparison with other embedding techniques. Section~\ref{sec:model-architecture} describes the proposed model. Section~\ref{sec:val} presents the results of evaluation. Section~\ref{sec:conc} concludes the work with discussion on future work.

\section{Related Work}
\label{sec:related}
There exist several comprehensive literature reviews on code clone detection which provide a great insight on the trend in the area. Ratten et al.~\cite{rattan2013software} reviewed the papers published until 2012 on Type I-IV clones and approaches to detect them. Azeem et al.~\cite{azeem2019machine} conducted a survey of the papers published from 2000 to 2017 that covers how machine learning is used to detect code smells including duplicate code that can be similar to code clone. Maggie et al.~\cite{lei2022deep} added to the previous by including up-to-date deep learning approaches on both single-language and cross-language clone detection where they found only a little work existing on the latter and emphasized the need for more research effort in the area. Sobrinho et al.~\cite{de2018systematic} covered types of sub-optimal code or bad smells that may lead to undesirable effects on code maintenance, where bad smells can be any of duplicate code (clone) and large class. 

Wei and Li~\cite{wang2020detecting} used LSTM to learn representations of code fragments in clone detection in a single language. LSTM uses seq2seq to learn output sequence from input sequence. LSTM autoencoder~\cite{said2020network, sutskever2014sequence} reconstructs output from decoded input one step at a time, which makes it suitable for sequence generation. The main disadvantage of seq2seq-based LSTM is their bottleneck issues as information tends to get lost, which affects the prediction capability of the model~\cite{raff2018malware, soyalp2021improving}. LSTM may be applied to cross-language clone detection where it can produce next AST tokens in one language for given AST tokens encoded in another language. We followed the same approach in our LSTM experiment but over InferCode embeddings.

Perez et al.~\cite{perez2019cross} used a 45k dataset of Java and Python clones. A vocabulary map is generated from a set of AST trees of programs, containing a set of tokens together with their types and values. Then, a skip-gram model is used to generate a token-level node vector representation where nodes belong to an AST of source code in Python or Java. These token-level vectors of nodes are then fed as input to a supervised neural model that learns the embeddings of the vectors so that it becomes easier to compare two probable clones based on their embedding. They take the advantage of AST structures to define the context of a token in AST instead of windows used in natural language processing. A node context is customizable in the skip-gram model to include as many children as possible. The context also includes one parent, which establishes a child-parent relationship in the context. The context of a node is defined by the ancestors, descendants, and siblings of the node within a given window which is amenable. The skip-gram model used in their work is very similar to Tree-based Convolutional Neural Network (TBCNN)~\cite{mou2016convolutional, mou2015discriminative}. The later, however, uses a context window of size 2. They also used three additional algorithms to find the context of a node and produced a list of node indices along with node contexts. Once data generation is done, the data is used to train the skip-gram model using the negative sampling objective loss function. Finally, the vectors learned through the skip-gram model are passed to a Siamese network to find similarity score between two code snippets written in Java and Python. They found that increasing the sibling window decreases clone detection, while keeping the window size short for ancestors and descendants increases clone detection accuracy, which is confirmed with TBCNN results.  

Sajnani et al.~\cite{sajnani2016sourcerercc} developed a SourcererCC that is token-based large-scale clone detection tool based on the lexical approach~\cite{nishi2018scalable}, which has the ability to achieve high precision and recall as compared to other tools and works best with near-miss Type III clone (not Type IV though), where Type III differ at statement level and near miss to mean that minor to significant editing changes occur in the copy/pasted fragments. They included 25k projects with 250 millions of code and 3 million examples. The tool works by comparing bag-of-tokens of code fragments that uses an optimized partial index of sub-blocks of a code block to quickly query the potential clones to achieve high scalability. The code block is reduced to $(token, frequency)$ pairs, where $frequency$ denotes the number of times token appeared in a code block. A token could be a keyword, literal, and identifier from a given source code. They use overlap similarity function between two pairs of source code. Overlap has to achieve certain threshold $\alpha$ to be considered as a clone.

Saini et al.~\cite{saini2018oreo} put a method-level, named Oreo, to detect clones in the twilight zone that is between Type III and Type IV, which is very difficult to detect. Oreo combines machine learning, information retrieval, and lexical-based software metrics,  semantic scanner on methods to find clones. BigCloneBench dataset has a variety of clone types between Type III and Type IV including Very Strongly Type-3, Strongly Type-3, Moderately Type-3, and Weakly Type-3~\cite{svajlenko2016bigcloneeval}. Each method in a pair has to go through a combination of size-based and semantic-based filters in order to survive and reaches machine learning classifier for clone detection. Sized-based filter mandates that a pair of methods sizes should be close. Semantic-based filter based on action tokens extracted from methods calls in the code body and fields, so for two methods they find the overlap between tokens $\left<t, freq\right>$ to find if they are semantically similar, where $t$ is the action token and $freq$ is the frequency of action token~\cite{goffi2014search}. Then the pair is passed to another a supervised machine learning based on Siamese architecture to assist further in detecting if the pair is clone. Out of 24 software metrics they used to filter out non-probable clones, 5 are proposed by Saini et al. which are type literals as they noticed that methods with similar functionality have similar literal types. They also reported that action tokens not only good for semantic comparison but also in detecting Type III clone~\cite{sajnani2016sourcerercc}. Software metrics are resilient to changes in identifiers and literals, which make them very popular not only to detect Type I and Type II clones~\cite{kontogiannis1997evaluation, mayrand1996experiment, patenaude1999extending} but also up to Type III clones. where for a given pair of methods, they should have close software metrics to be considered as clones based on a predefined threshold $\lambda$ decided by author. After collecting metrics from data, a vector of size 48 is fed to Siamese architecture with dropout, relative entropy loss function and stochastic gradient descent optimizer.  

Nafi et al.~\cite{nafi2019clcdsa} proposed the CLCDSA model to detect cross-language code clones using both semantic features and syntactic features of pairs of code fragments in Java, Python, and C\#. After heuristic and manual study, they elicited 9 features out of 24 total features from Saini et al.'s work~\cite{saini2018oreo} as based on heuritsitc study, they found these 9 features are more frequent in cross-language setup. Moreover, these 9 features yield similar metrics between different but functionally similar programming languages. While their model achieved higher scores compared to Perez et al.'s work~\cite{perez2019cross}, there are many preconditions imposed. For example, in addition to exhaustive feature selection, they included API documentation as part of finding clones, which adds an extra burden on the CL model. Also, they demand that pairs have to go through the API call similarity filter based on the skip-gram model to measure the possibility of being clones before clone detection starts. To conclude, this is not an end-to-end approach and extensive feature processing has to take place ahead. 

\section{Source Code Embeddings and InferCode}
\label{sec:background}
In this section, we give an overview of embedding techniques and  InferCode~\cite{bui2021infercode} which is used as the main source code embeddings in this work. There are many source code embedding techniques published recently including code2seq~\cite{alon2018code2seq} and code2vec~\cite{alon2019code2vec} which are denoted as code2*. These approaches enumerate over a set of AST $k$ paths between terminals. As a result, a pair of terminals is selected each time different from all other pairs. It then uses an attention mechanism to select the path that will most likely contribute to the prediction. Strictly speaking, their embeddings are meant only for a specific task in mind, which can threat the validity of our work if we are to use them for cross-language clone detection. Subsequent attempts improved code2* by obfuscating code which increased accuracy for many downstream tasks including method name prediction~\cite{compton2020embedding}. code2* take AST paths as input and fed them to a recurrent neural network (RNN) powered with attention to recognize which subtree contributes the most to predict the label. code2* encode all AST paths and then determine which path contributes the most to the final output during decoding. At the decoding step, code2* run over each path from AST and decode it to the target sequence one token at a time using Long Short Term Memory (LSTM)~\cite{hochreiter1997long}. The decoder also uses an attention mechanism with decoder hidden state to produce each target token one at a time.  

\begin{figure}[!ht]
    \centering
    \includegraphics[scale=0.28]{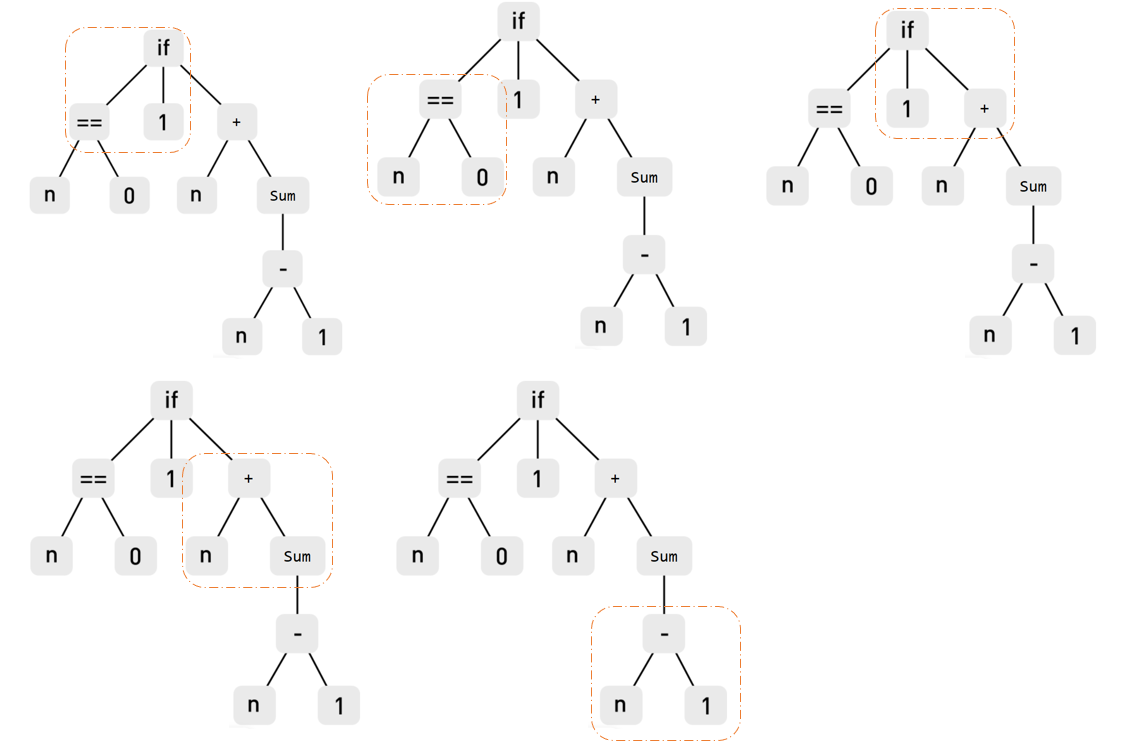}
    \caption{TBCNN loops over a subtree for embedding. Each parent node accumulates the information of its descendants.}
    \label{fig:tree_cnn}
\end{figure}

InferCode is an end-to-end self-supervised model that does not need a human to label data, which boosters the model's generalizability. The model also learns labels from the code by itself. InferCode uses TBCNN to loop over AST to embed the whole subtree into a vector that describes the whole code snippet, so that it becomes easier to compare the given code to other source code snippets in different languages. Fig.~\ref{fig:tree_cnn} shows looping over an subtree. Each node in the subtree $v_i$ in AST tree $T$ is associated with $D$-dimensional vector $x_v \in \mathbb{R}^D$ as input feature that represents $x_v=x_{type}+x_{token}$ (token and text of node $v_i$) and then hidden representation $h_v$ is extracted. Embeddings for types and texts of nodes are also learnable as $W^{type}, W^{text}$. Also, 3 additional weight matrices are introduced for each node $W^t, W^l, W^r$ representing top, left and right. 

InferCode tweaked TBCNN to include the textual information of code besides the type information provided by the original implementation of TBCNN. InferCode also replaces the dynamic max pooling layer with an attention mechanism to reduce information loss in the aggregation of all information from all nodes into a single vector where each parent accumulates the information of its children nodes at one level. 
InferCode utilizes a special filter that is designed to loop over subtrees and embed their information to output a feature map that replaces the traditional convolution filter used in convolutional neural networks (CNNs).
Based on the subtree of source code AST, InferCode predicts the possibility of a subtree belonging to an AST. 


\section{Model Architecture}
\label{sec:model-architecture}
In this section, we describe how we used Siamese architecture to take pairs of code snippets embeddings to find out if they are clones or not. Unlike Perez et al. work, we did the opposite by using shared weights between the 2 branches of the network because we think that existing embeddding for both are based on statement level as we explained how TBCNN work, so we consider both to be very close to each other and in the same domain.

\subsection{Subtrees Comparison}
It is a good start to loop over subtrees of AST, but this leads to an exponential running algorithm. For two binary trees, which is not the case mostly for AST of code, of levels $i,j$ respectively, we have to loop over $2^i \times 2^j$ in order to find out if the two ASTs for two source codes written in Python and Java are similar. We can add a margin on how many subtrees has to match before we consider two ASTs to be similar or clones. What is more interesting is if can aggregate all information obtained from all nodes before the a specific node in AST that will yield the context around that node. We found that InferCode built on similar reasoning~\cite{bui2021infercode}. This not only helps to detect exact clones but also near miss clones. 

\begin{figure}[!ht]
    \centering
    \includegraphics[scale=0.48]{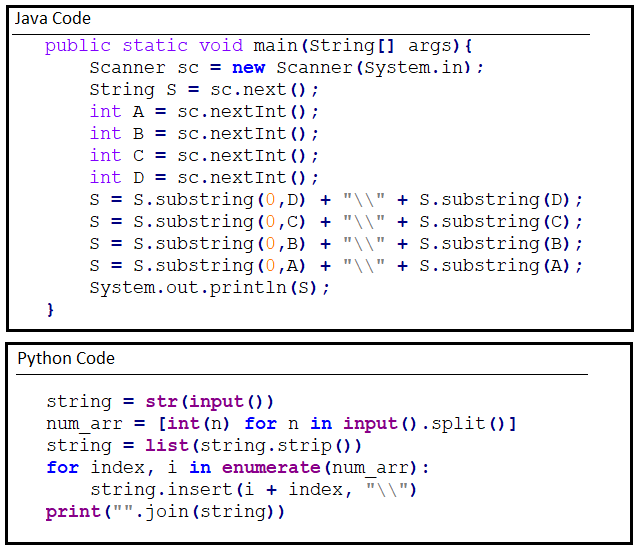}
    \caption{Python and Java code that iterates over substrings and add a backslash `\textbackslash' between all substrings before appending all substrings back together. This might better fit Type IV clone category.}
    \label{fig:java_py_clone}
\end{figure}

To reduce number of subtrees comparison for cross CL, we can use hashing trick by hashing all possible subtrees of source codes in Python and Java and then compare those that fits in the same bucket~\cite{baxter1998clone}. This can be challenging if there is a slight variations among subtrees as slight change in subtree will yield different bucket in the hash table, so mainly depending on exact subtrees matching into our approach is not wise as Python and Java ASTs does not share much similarity at syntax level compared to code clones written in the same language. Consider the following clones in Python and Java that are supposed to take substrings of input string and append all substrings together~Fig.\ref{fig:java_py_clone}. Both subtrees and tokens names taken would help in previous scenario scenario as if we take token names or subtrees individually to detect clones would be difficult. If the programs were written in the same language, we would have both semantic and syntactic match~\cite{bulychev2008duplicate, jiang2007deckard}, also it can be the case for cross language clones.

We should also do the same across all other languages in case the same functionality is replicated. Fig.~\ref{fig:java_py_clone} is one simple example where clone detection can prove helpful, although it can be more intricate in production software as fragments pieces can have complex functionalities.

In Fig.\ref{fig:java_py_clone}, although the two programs are functionally the same, yet their API call similarity is quite different. Java API calls made are IN(), SUBSTRING() and NEXTINT() and for Python API calls are INPUT(), SPLIT(), STRIP() and INSERT(). Only INPUT and IN are similar both at syntactic level and API documentation level, however the others are different. So, API similarity action filter proposed by CLCDSA will drop our example provided above. 

\subsection{Siamese Architecture}
\label{sec:SiameseArcheticture}
Simple cosine similarity metrics or Euclidean distance metrics can be used to find how similar two embeddings are. However, we decided to use Siamese architecture for this purpose as we are looking to train the network to learn similarity of embeddings pairs and also dissimilar pairs. Siamese architecture is an architecture built exclusively for similarity computation. It is composed of two or more equivalent parallel networks that have the same architecture and trainable parameters~\cite{bromley1993signature}. Parallel networks are updated during backpropagation with the same value. In addition, Siamese architecture needs less data to be trained, which makes it suitable in this work as we deal with more than $40k$ pairs of similar and dissimilar embeddings of Java and Python code. Siamese architecture has been widely used since it was developed for various tasks such as images similarity, text similarity, plagiarism detection, and many others (e.g., \cite{chopra2005learning, ichida2018measuring, ranasinghe2019semantic}).

\begin{figure}[!ht]
    \centering
    \includegraphics[scale=0.45, angle =-90]{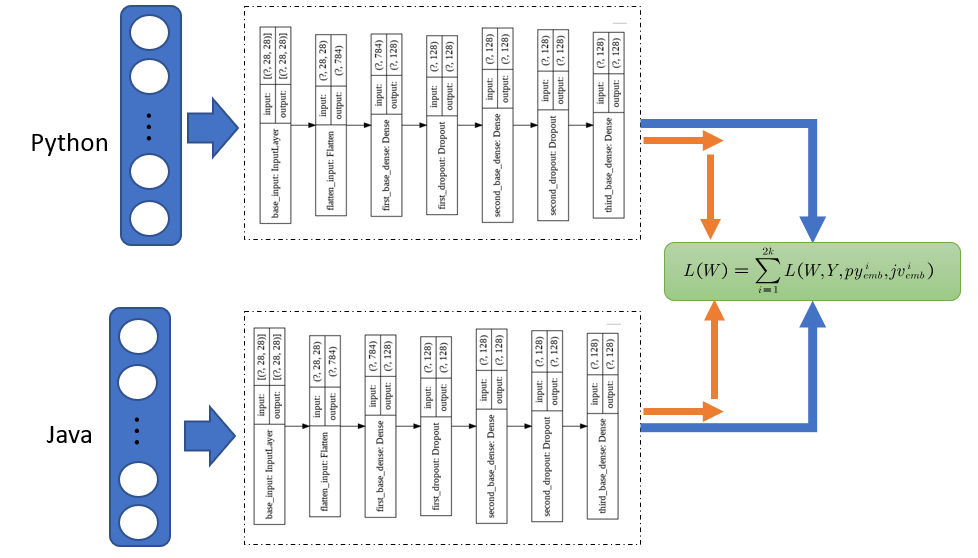}
    \caption{Siamese model we used along with 2 parallel networks one for each input of a pair whether it is clone or not.}
    \label{fig:Siam}
\end{figure}

We provide our Siamese architecture with pairs of similar embeddings followed by dissimilar embeddings. We annotate the dataset with similarity score in $label\in\{0, 1\}$ or a label that explicitly tells that a pair is similar or not. We also use the contrastive loss function to calculate the distance of every pair and find how far they are based on the preset threshold~\cite{bromley1993signature}. Fig~\ref{fig:Siam} shows two sub models to process the pairs which can be similar and dissimilar. Then, we measure the similarity between the output of the two networks using the Euclidean distance. During the backpropagation of the model per the output of the loss function, the layers are updated based on the similarity or dissimilarity of pairs. 

Although Euclidean distance is suitable for measuring the distance between two points, in a large dimensional space, all points tend to be far apart by the Euclidean measure. In higher dimensions, the angle between vectors is a more effective measure. The cosine distance measures the cosine of the angle between the vectors. The cosine of identical vectors is 1 while orthogonal and opposite vectors are 0 and -1 respectively.

The contrastive loss function is proven to be effective in unsupervised learning. This loss will learn from neighbourhood embeddings of source code by pushing similar pairs together and dissimilar pairs apart, which makes it suitable in our use case. Contrastive loss looks similar to the softmax function with the addition of the vector similarity and a normalization factor. The similarity function is just the cosine distance discussed earlier. The other difference is that values in the denominator are the cosine distance from the positive samples to the negative samples; not very different from CrossEntropyLoss~\cite{zhang2018generalized}.
The intuition here is that we want our similar vectors to be as close to 1 as possible, since that is the optimal loss. We want the negative samples to be close to 0, since any non-zero values will reduce the value of similar vectors. Given embeddings of Python and Java pairs (clones or not clones) defined as $\forall py_{emb}, java_{emb} \in \mathbb{R}^N$, where $N$ is the dimension of the embedding, the task is the find the distance between the output $O_W(emb)$ of Siamese
network parameterized by weight matrix $W$:
\begin{equation}
  \begin{aligned}
  & D_W\left( py_{emb},java_{emb} \right) =\parallel O_W\left( py_{emb} \right) \\  & -O_W\left( java_{emb} \right) \parallel _2
  \end{aligned}
\end{equation}

Once that is found, the output is passed to the loss function for one pair:
\begin{multline}
L( Y,W,py_{emb},jv_{emb} ) =( 1-Y ) \frac{1}{2}( D_W)^2 + \\ 
( Y ) \frac{1}{2}\{ \max ( 0,\ m-D_W ) \} ^2
\end{multline}
where $Y \in \{0,1\}$, where $1$ indicates the pair is clone and $0$ otherwise. To find the total loss of all pairs $2K$, as we have $K$ clone pairs and $K$ dissimilar pairs:
\begin{equation}
  \begin{aligned}
L( W ) =\sum_{i=1}^{2k}{L( W,Y,py_{emb}^{i},jv_{emb}^{i} )}
  \end{aligned}
\end{equation}

For similar pairs, we have $\left( 1-Y \right) \frac{1}{2}\left( D_W\left( x_i \right) ^2 \right) +0$ and $0+\left( Y \right) \frac{1}{2}\left( \max \left\{ 0,\ m-D_W\left( x_i \right) \right\} ^2 \right) $ when the pairs are dissimilar, where $x_i$ is embedding $i$. To penalize the loss function, we need to have a margin $m$ when our inputs are dissimilar. If distance goes above this margin $m<D_W\left( x_i \right)$, it will yield a negative number and thus, 0 will be chosen. In that case, the whole equation will equate to 0 if we exceed the margin and the network is not updated. 

\section{Evaluation}
\label{sec:val}
In this section, we evaluate two models on the dataset of clones and dissimilar pairs. The first model is LSTM with teacher forcing and Bahdanau Attention over AST tokens. The second is our model over the same dataset. The trained model used in this work and its manual are available online~\footnote{https://bitbucket.org/MohammadAbrahiam/crosslanguagecloneyahyaetal/src/master/}. 

\subsection{LSTM with Bahdanau Attention}
We use LSTM with the Bahdanau Attention formulas~\cite{chorowski2015attention} to find attention. We have formulated the cross language clone detection task in our early experiments as a sequence to sequence (seq2seq) problem where we take ASTs of code snippets as input in one programming language and feed it to the LSTM autoencoder. The encoder of LSTM takes the parsed ASTM tokens one by one at each timestep. The decoder then will decode output one step at a time. Then we challenged LSTM based seq2seq model to map input in of input AST to output AST. The model achieved 0.5 F1 score, but with poor generalization in Fig. below. While passing hidden states between LSTM unit cells as time goes increases accuracy as results below show, not passing states between layers does not perform poorly. The goal is to encode Python script and then decode to get the clone in Java. We will use as well in this experiment teacher forcing to accelerate training of LSTM Autoencoder. We have trained the LSTM model with 16 unit cells, train data 14280, test size 3569, 500 epochs, 20 patience degree, adaptable learning rate, $rmsprop$ optimizer and $mse$ loss. We tried different testing scenarios by changing number unit cells and optimization function, but the model stops between 22-27 epochs with negligible increase in F1 score. F1 score for this model 0.53 F1 and 0.31 validation loss. 

\begin{figure}[!ht]
    \centering
    \includegraphics[scale=0.25]{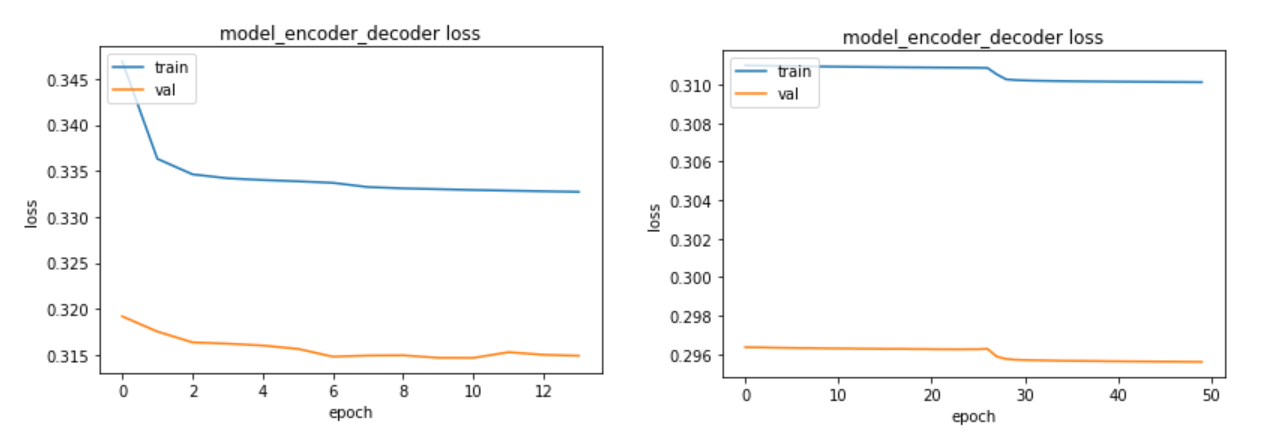}
    \caption{Training loss (blue) and validation loss (orange) of LSTM+TF on the right and LSTM+BL on the left. The training loss of LSTM+BL converges to validation loss better than LSTM+TF.}
    \label{fig:lstmbftf}
\end{figure}

In teacher forcing, we use the ground truth output in the current time step (available in the training data) to compute the system state in the next time steps. Teacher Forcing is a method to train encoder-decoder models in Seq2Seq model to accelerate training. Teacher Forcing can only be used at Training as experiments show we get bad predictions. Even though Teacher Forcing improves the training process by fast converging, the model has generate low accuracy even with the training data. Nevertheless, LSTM model with TF gives better F1 score than LSTM with BL despite that LSTM with BL shows stabilized training as Fig.~\ref{fig:lstmbftf} shows.

Problem with vanilla LSTM Autoencoder model is the use of a fixed-length context vector. Dzmitry et al. conjectured that this limitation may make the basic encoder–decoder approach to underperform with long sequences and as a result information will be lost. To verify this conjecture, we conducted experiments on different time steps of LSTM model: 1)When the sequence size or number of time steps is 4, encoder-decoder model terminates at Epoch 31 with \%99 accuracy score, and 2)When the number of time steps is 16, encoder-decoder model runs all the 40 epochs and finishes with only \%36 accuracy score. That is, as argued, Encoder Decoder model underperforms with long sequences. We use not only the last hidden and cell states but also the decoder’s hidden states generated at all the time steps. Also, we use all the decoder’s hidden states at all consecutive time steps. We start by initializing the decoder states by using the last states of the encoder as usual. Then at each decoding time step: 1) We use encoder’s all hidden states and the previous decoder’s output to calculate a global context vector by applying an Attention Mechanism, and 2) Lastly, we concatenate the Context Vector with the previous Decoder’s output to create the input to the decoder. We used in LSTM+BL (LSTM with Bahdanau Additive style) experiment $v_{a}^{T}\tan\text{h}\left( W_1h_t+W_2h_s \right)$ score as defined by BL Style~\cite{chorowski2015attention},

\begin{multline}
 score\left( h_t,h_s \right)=\begin{cases} 
      h_{t}^{T}Wh_s \\
      v_{a}^{T}\tan\text{h}\left( W_1h_t+W_2h_s \right)
   \end{cases}
\end{multline}

where $h_s$ represents all the hidden states of the encoder, $h_t$ is the previous hidden states of the decoder (previous time step output), and finally $W$ is the weight matrix for parameterizing the calculations. To find attention (which AST tokens are most likely to influence mapping Python code to Java code), we calculate a score to relate the Encoder’s all hidden states and the previous Decoder’s output. So we will compare encoders all hidden states and previous decoder’s output to create a score. Then calculate attention weights is very similar a softmax of the values we calculated in context vector step,
\begin{equation}
  \begin{aligned}
\alpha _{ts}=\frac{\exp \left( score\left( h_t,h_s \right) \right)}{\sum\limits_{s'=1}^S{\exp \left( score\left( h_t,h_s \right) \right)}}\ \ \
  \end{aligned}
\end{equation}
Finally, to calculate the context vector $c_t$ below by applying the attention weights onto decoder hidden states $h_s$. Thus, we will have weighted decoder hidden states at the end. This will weaken features from encoder hidden states that attribute less to current decoder output hidden state,
\begin{equation}
  \begin{aligned}
c_t=\sum_s{\alpha _{ts}h_s}
  \end{aligned}
\end{equation}

\begin{table}
\centering
\footnotesize 
\caption{LSTM Model Results on Cross Language Clone detection between Java and Python embeddings. LSTM AE stands for LSTM Autoencoder. LSTM + TF stands for LSTM with teacher forcing. LSTM with BL stands for Bahdanau and Luong Attention.}
\label{LSTMCrossLanClone}
 \begin{tabular}{|l|l|l|l|}
\hline
\textbf{Models} & \textbf{Precesion} & \textbf{Recall} & \textbf{F1-Measure}\\
\hline 
Vanilla LSTM AE& 0.60 & 0.45 & 0.53\\
\hline 
LSTM+TF & 0.62 & 0.53 & 0.56\\ 
\hline
LSTM+BL  & 0.59 & 0.45 & 0.54\\
\hline 
\end{tabular}
\end{table}

\subsection{Siamese Model Network}
Siamese network accepts pairs from the same domain. Unlike Perez et al.'s work, they did not share the weights across base network of Siamese  network for the pair as they considered them to be in different domains as one source code is in Java while the other is in Python. Instead, we did the opposite and we share the weights across the two branches of base network that accept pair. 

Based on experiments, we noticed that Siamese networks perform better in terms of F1 score than other loss functions when it's used with contrastive loss. In our network, SGD optimizer with patience level set to 20 for early stoppings to avoid overfitting. There are many version of contrastive loss function, but the one that we used as stated in~\ref{sec:SiameseArcheticture} is quadratic contrastive loss function. For the layers, a set of dense and dropout layers used alternately. 

\begin{table}
\centering
\footnotesize 
\caption{Siamese Model Results on Cross Language Clone detection between Java and Python embeddings compared to other models.}
\label{LSTMCrossLanClone}
 \begin{tabular}{|l|l|l|l|}
\hline
\textbf{Models} & \textbf{Precision} & \textbf{Recall} & \textbf{F1-Measure}\\
\hline 
Our Model & \textbf{0.99 } & 0.59 & \textbf{0.80}\\
\hline 
CLDSA & 0.67 & 0.65 & 0.66\\ 
\hline
Daniel's Siamese Model & 0.66 & 0.83 & 0.66\\ 
\hline 
\end{tabular}
\end{table}

\section{Conclusion}
\label{sec:conc}
While same language clone detection achieves very high accuracy, cross language clone detection is still ongoing to improve tools accuracy to deal with multiple types of clones across many languages. The difficulty in cross language clone detection task lies in the fact that both have different structure and statements  In this work, we demonstrated how Siamese architecture can be used to learn more abstract embedding from source code embeddings obtained over AST across different programming languages. It's good to notice that no labeling is need need for our similar and dissimilar pairs. We added to the dataset over 20k pairs of dissimilar clones besides 20k pairs of clones, which is very important for Siamese architecture to learn similarity between pairs. So, total we have 80k rows in our dataset. We showed that our model improves the state-of-the-art work by over \%10. It's true that our model works best with Type III clones, but we are looking to further test the model on Type IV clones and increase the accuracy of the model on Type III clone as well across different languages. 

\bibliographystyle{ACM-Reference-Format}
\bibliography{Ref}

\end{document}